\documentclass[epj,final]{svjour}
\usepackage[dvips]{graphicx}
\usepackage[dvips,usenames]{color}
\usepackage{amssymb,amsmath}
\usepackage{float}
\usepackage{appendix}

\providecommand{\Prob}{\text{Prob}}
\providecommand{\openone}{\leavevmode\hbox{\small1\kern-3.8pt\normalsize1}}

\newcommand{\avk}{\langle k \rangle}

\newcommand{\al}{\alpha}
\newcommand{\av}[1]{\langle {#1} \rangle}

\begin{document}

\title{Steady-State Dynamics of the Forest Fire Model on Complex
  Networks} \author{Jean-Daniel Bancal\inst{1} \and Romualdo
  Pastor-Satorras\inst{2}}

\institute{Institut de Th\'eorie des Ph\'enom\`enes Physiques, Ecole
  Polytechnique F\'ed\'erale de Lausanne, CH-1015 Lausanne,
  Switzerland \email{jean-daniel.bancal@unige.ch} \and Departament de F\'isica
  i Enginyeria Nuclear, Universitat Polit\'ecnica de Catalunya, Campus
  Nord, 08034 Barcelona, Spain}

\date{\today}

\abstract{Many sociological networks, as well as biological and
  technological ones, can be represented in terms of complex networks
  with a heterogeneous connectivity pattern. Dynamical processes
  taking place on top of them can be very much influenced by this
  topological fact. In this paper we consider a paradigmatic model of
  non-equilibrium dynamics, namely the forest fire model, whose
  relevance lies in its capacity to represent several epidemic
  processes in a general parametrization. We study the behavior of
  this model in complex networks by developing the corresponding
  heterogeneous mean-field theory and solving it in its steady
  state. We provide exact and approximate expressions for homogeneous
  networks and several instances of heterogeneous networks.  A
  comparison of our analytical results with extensive numerical
  simulations allows to draw the region of the parameter space in
  which heterogeneous mean-field theory provides an accurate
  description of the dynamics, and enlights the limits of validity of
  the mean-field theory in situations where dynamical correlations
  become important.  \PACS{ {PACS-key}{discribing text of that key}
    \and {89.75.-k}{Complex systems}\and {64.60.aq}{Networks}
  } 
   } 

\authorrunning{J.-D. Bancal and R.  Pastor-Satorras}
\titlerunning{Steady-State Dynamics of the FFM on Complex
  Networks}
\maketitle
\section{Introduction}
\label{intro}

The heterogeneous topology of a networked substrate has been proven to
have a large impact on dynamical processes taking place on top of it
\cite{barrat08:_dynam,dorogovtsev07:_critic_phenom}. These topological
effects are especially remarkable in the case of scale-free (SF)
complex networks \cite{BA,caldarelli2007sfn}, characterized by a
degree distribution $P(k)$, defined as the probability that an element
in the network (vertex) is connected to $k$ other elements, that exhibits a
power-law behavior, $P(k) \sim k^{-2-\gamma}$, with $0 < \gamma \leq
1$ \footnote{To ease the notation in our mathematical treatment, we
  will use this definition of the degree exponent in a power-law
  degree distribution.}. The diverging second moment $\av{k^2}$ of the
degree distribution has thus been found to be at the core of the
peculiar behavior observed in a wide array of non-equilibrium dynamical
processes, ranging from percolation \cite{newman00,havlin01},
absorbing-state phase transitions \cite{castellano06:_non_mean},
self-organized criticality
\cite{PhysRevLett.91.148701,0295-5075-57-5-765}, synchronization
phenomena \cite{arenas08:_synch}, opinion dynamics
\cite{castellano2007sps}, etc.

The interplay between topology and dynamics has been particularly
studied in the case of epidemic processes \cite{keeling05:_networ},
where the relevant substrate is the network of contacts through which
the disease spreads \cite{amaral01}. Starting from the first
observations of an epidemic threshold scaling as the inverse of the
second moment of the degree distribution, and thus vanishing in the
thermodynamic limit of an infinite network size
\cite{Romualdo:2001,lloyd01,moreno02,newman02b}, a wealth of
interesting and relevant results have arisen, dealing, to mention just
a few, with immunization strategies \cite{psvpro,cohen-2003-91},
effects of bipartite (heterosexual) populations
\cite{gardenes08:_spread} or epidemic forecasting
\cite{colizza06:_predic}.

The understanding of the features of epidemic spreading is mainly
based on the analysis of compartmental models \cite{anderson92}, in
which the population is divided into different classes, according to
the stage of the disease. Individuals (the vertices in the network)
are in this way classified as susceptible (healthy and capable to
contract the disease), infected (sick, and capable to transmit the
disease), recovered (immunized or dead), etc. With these definitions,
different epidemic models can be formulated, according to the
succession of states that the evolution of the disease imposes on each
individual, such as susceptible-infected-susceptible (SIS),
susceptible-infected-recovered (SIR),
su\-sc\-ept\-ib\-le-infected-removed-su\-sc\-ep\-tible (SIRS), etc.
The theoretical analysis of the behavior of these compartmental models
in complex networks starts from the application of the heterogeneous
mean-field (HMF) theory
\cite{barrat08:_dynam,dorogovtsev07:_critic_phenom}.  This formalism
is based on the assumptions that all vertices with the same number of
connections (i.e.  within the same degree class) share the same
dynamical properties, and that fluctuations are not important, and
therefore all relevant variables can be described in terms of
deterministic rate equations. The first assumption becomes natural
once we admit that the degree is the only parameter describing the
state of a vertex. On the other hand, the second assumption finds
support in the small-world property shown by most complex networks
\cite{WS}, implying that dynamical fluctuations take place so close
together that they can be washed away in very few time
steps\footnote{As a matter of fact, fluctuations can be shown to be
  irrelevant in some particular cases \cite{boguna09}.}.  HMF has
proved to be extremely useful in providing an accurate description of
epidemic models on complex networks, and has in fact become the
\textit{de facto} standard tool to analyze general non-equilibrium
processes on such substrates \cite{barrat08:_dynam}.

In this paper we will pay attention on a non-equilibrium dynamical
model with relevance both in epidemic modeling and other ambits of
non-equilibrium statistical physics, namely the forest fire model
(FFM).  First introduced in 1992 by Bak \textit{et al.}
\cite{BakChenTang}, and further developed by Drossel and Schwabl
\cite{DrosselSchwabl}, the FFM was elaborated to show self-organized
criticality and avalanche behavior in a specific limit of its defining
parameters.  Even though its general status as a self-organized
critical model is under debate
\cite{ISI:000174484100001,bonachela09:_self}, it has found successful
applications as a general disease propagation model
\cite{keeling05:_networ,rhodes96:_power_laws}, in which susceptible
individuals can get the disease either by transmission from an
infected neighbor or spontaneously (because of a mosquito bite for
instance), while recovered individuals can become again
susceptible. It is thus akin to a SIRS model \cite{keeling05:_networ}
with an external source of infected individuals. More interestingly,
it encompasses several other compartmental models, which can be
recovered in a convenient way as certain limits of the parameters that
define the FFM.

Previous works on the FFM in complex networks have reported, among
other results, the presence of self-sustained oscillations in
small-world networks \cite{SIRS} and analyzed the distribution of
excitations depending on topology \cite{MullerLinowMarrHutt}.  From
the perspective of the SIRS model, on the other hand, its
epidemiological implications have been discussed for certain ranges of
its parameter space \cite{liu04:_spread,4593477}.  In the present
paper we provide an extensive theoretical analysis of the FFM, using
the HMF formalism and focusing on the steady-state dynamics of the
model in the whole range of its parameter space. Our analysis allows
to emphasize its interpretation, in the different regimes, in terms of
known epidemic models, providing analytic expressions for the
steady-state density of infected individuals in certain limits of the
relevant parameters. A comparison of the theoretical results with
extensive numerical simulations, allows finally to unveil the
limitations of the HMF approach in this and probably other epidemic
models, hinting towards the break down of HMF theory when dynamical
correlations become relevant \cite{boguna09}.

We organized our paper as follows: In Sec.~\ref{sec:1} we define
the FFM, discussing its relation to self-organized criticality and
disease propagation. In Sec.~\ref{mean-field} we develop the HMF
theory of the FFM in general complex networks. Sec.~\ref{sec:2} deals
with the steady-state solution of the HFM equations obtained
before. We show in particular how an appropriate rescaling of the
equations allows to simplify the description and to conveniently
reduce the number of degrees of freedom. A general analysis is
presented for both homogeneous networks and heterogeneous networks
with no spontaneous infection. An explicit analysis of uncorrelated SF
networks is presented in Sec.~\ref{sec:3}. The numerical simulations
shown in Sec.~\ref{sec:4} allow us to check the validity of our
theoretical results, as well as to draw the limits of validity of
general HMF approaches. Finally, we present our conclusions in
Sec.~\ref{concl}.

\section{Forest fire model on complex networks}
\label{sec:1}

We consider the FFM on general complex networks which, from a
statistical point of view, are described at a coarse-grained level by
the degree distribution $P(k)$ and the degree-degree correlations,
given by the conditional probability $P(k'|k)$ that a vertex of degree
$k$ is connected to a vertex of degree $k'$ \cite{alexei}.

\begin{figure}[t]
\begin{center}
\includegraphics[width=0.18\textwidth]{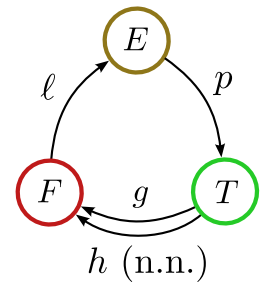}
\end{center}
\caption{Representation of the three states present in the forest fire
  model, and of its dynamical rules. (n.n. stands for a ``nearest
  neighbor'' interaction).}
\label{fig:ETF}
\end{figure}

In the FFM, each vertex in the network is in one of three excluding
states: E (empty), T (tree), F (burning tree). The evolution of the
model is defined in a continuous time formulation in terms of the
possible events that can happen in a small time interval $\Delta t$
(see Fig.~\ref{fig:ETF})
\begin{enumerate}
\item{E $\to$ T : } A tree can grow on an empty vertex with
  probability $p \Delta t$.
\item{T $\to$ F : } For a tree, each of its burning neighbors
  (if any) can light it with probability $h\Delta t$. 
\item{T $\to$ F : } Additionally, there is also a probability $g
  \Delta t$ for a tree to catch fire spontaneously (e.g. mediating a
  lightning). These two burning events are considered
  probabilistically independent.
\item{F $\to$ E : } A burning tree leaves an empty vertex with
  probability $\ell \Delta t$.
\end{enumerate}

The FFM was proposed to exhibit self-organized criticality in the
double limit $p\to0$ and $g/p \to0$, in which clusters of trees are
allowed to grow before burning down, leading to a distribution of fire
avalanches with an approximately power-law form \cite{jensen98}. For
finite $p$ and $g$ values, an activated dynamics with no evidence of
avalanches is instead observed \cite{SIRS,MullerLinowMarrHutt}.

From the point of view of disease modeling, trees, fires, and empty
sites represent, respectively, susceptible, infected, and recovered
individuals. Within this interpretation of the model, susceptible
individuals can acquire the disease by contact with one or more
infected individuals at rate $h$; infected individuals recover in a
time scale of the order $\sim 1/\ell$; recovered individuals become
again susceptible in a time scale $\sim 1/p$; and healthy individuals
become spontaneously infected in a time scale $\sim 1/g$. With a set
of $4$ parameters, a unified description of the most usual epidemic
models is achieved. Thus, in the limit $p=g=0$, we recover the SIR
model; the limit $g=0$ and $p\to\infty$ leads to the SIS model; while
the limit $g=0$ corresponds to the SIRS model.

Note finally that our definition of the FFM considers time as a
continuous variable, in opposition to previous approaches. This
formulation is preferred in order to lead more naturally to a
continuous analytical description in terms of differential equations,
and will be taken into account when performing numerical simulations
in Sec.~\ref{sec:4}.

\section{Heterogeneous Mean-Field theory for the FFM in complex
  networks}
\label{mean-field}

Within the HMF approach, a dynamical system is assumed to be fully
determined in terms of the relative probabilities that a vertex of
given degree $k$ is in any one of the states allowed by the dynamics
\cite{Romualdo:2001,moreno02}. In the case of the FFM, this
description invokes the partial densities $\rho^\alpha_k(t)$, defined
as the conditional probability that a vertex of degree $k$ is, at time
$t$, in the state $\alpha$, with $\alpha \in \{E, T, F\}$. Since each
vertex must be in one of these states, the partial densities satisfy
the normalization condition
\begin{equation}
  \rho^E_k(t) + \rho^T_k(t) + \rho^F_k(t)=1.
\end{equation}
Therefore, only two independent partial densities, say $\rho^E_k(t)$
and $\rho^F_k(t)$, must be considered in the analysis. On the other
hand, the density of vertices in each state at time $t$ is given by
\begin{equation}
  \rho^\alpha(t) = \sum_k P(k) \rho^\al_k(t).
\end{equation}

At the core of the HMF theory lie the rate equations fulfilled by the
partial densities. By considering the different microscopic steps
allowed in the model we can readily write the change of the quantities
$\rho^\al_k(t)$ in an infinitesimal time step $\Delta t$, that are
given by
\begin{eqnarray}
  \rho^E_k(t+\Delta t)&=&\rho^E_k(t)+\rho^F_k(t) \Prob_k(F\to
  E) \nonumber\\
  &-& \rho^E_k(t)\Prob_k(E\to T), \label{eq:1}\\
  \rho^T_k(t+\Delta t)&=&\rho^T_k(t)+\rho^E_k(t) \Prob_k(E\to
  T) \nonumber\\
  &-& \rho^T_k(t)\Prob_k(T\to F),\label{eq:2}\\
  \rho^F_k(t+\Delta t)&=&\rho^F_k(t) - \rho^F_k(t) \Prob_k(E\to
  F)  \nonumber \\
  &+&\rho^E_k(t)\Prob_k(T\to F),\label{eq:3}
\end{eqnarray}
where $\Prob_k(\al\to\beta)$ is the probability that a vertex of
degree $k$ experiences the transition from the state $\al$ to the
state $\beta$ in a time interval $\Delta t$. From the definition of
the FFM in Sec.~\ref{sec:1}, we can immediately write down
$\Prob_k(E\to T)= p \Delta t$ and $\Prob_k(F\to E)= \ell \Delta t$. In
order to construct the term $\Prob_k(T\to F)$, we must consider that
rules 2 and 3 defining the model, which represent a tree catching
fire, are statistically independent. Therefore if we define $H$ as the
event ``A tree is lighted by its neighbors'' and $G$ as the event ``A
tree lights up spontaneously'', we have that $\Prob_k(T\to
F)=\Prob_k(G\cup H)=\Prob_k(H)[1-\Prob_k(G)]+\Prob_k(G)$. Rule 3 gives
$\Prob_k(G)= g \Delta t$. Now, since in this description vertices with
the same degree are statistically equivalent, the state of a given
vertex is independent on the state of its neighbors, and it only
depends on its degree. This allows us to write the probability for one
neighbor of a tree with degree $k$ to be burning as
\begin{equation}\label{eq:theta_k}
\theta_k=\sum_{k'}P(k'|k)\rho^F_{k'},
\end{equation}
given in terms of the average of the conditional probability $P(k'|k)$
that the vertex $k$ is connected to a vertex of degree $k'$, times the
probability that this last vertex is burning, $\rho^F_{k'}$. Notice that
here we are assuming that the edge through which $k'$ became burning
is immediately available to transmit again the fire. This assumption,
in opposition to the behavior of the SIR model
\cite{Epidemic_correlated_2}, will thus be valid only for $p >0$.

From rule 2, the probability that a particular nearest neighbor fire
ignites a tree in a vertex of degree $k$ is given by $h \Delta t
\theta_k$. Therefore, the probability that a tree of degree $k$ is
ignited by any of its nearest neighbors is $\Prob_k(H) =1 - [1-h
\Delta t \theta_k]^k$. Substituting this expressions in
Eqs.~(\ref{eq:1})-(\ref{eq:3}), and taking the limit $\Delta t \to 0$,
we obtain the final HMF equations for the FFM,
\begin{equation}\label{eq:mean_field}
  \begin{cases}
    \dot \rho^E_k(t)=\rho^F_k(t) -  p \rho^E_k(t)\\ 
    \dot \rho^T_k(t)=p \rho^E_k(t) -
    \rho^T_k(t)\left[h k\sum_{k'}P(k'|k)\rho^F_{k'}(t)+ g \right]\\ 
    \dot  \rho^F_k(t)=\rho^T_k(t)\left[h
      k\sum_{k'}P(k'|k)\rho^F_{k'}(t)+g \right]
    - \rho^F_k(t) 
  \end{cases},
\end{equation}
where we have set $\ell=1$, which amounts to a trivial rescaling of
time.  

Eqs.~(\ref{eq:mean_field}) represent a complete description of the FFM
at the HMF level. Even though we have derived them in a
phenomenological way \cite{Romualdo:2001,Endemic_states}, they can
also be obtained from a microscopic point of view, considering
explicitly the state of each vertex evolving as a Poisson random
process \cite{boguna09,Microscopic}. The mean-field result is then
recovered by averaging over the random processes and over the vertices
with same degree.

One final warning comment is in order here, concerning the fact that,
in writing Eqs.~(\ref{eq:mean_field}), we have neglected altogether
\textit{dynamical correlations} between adjacent vertices, assuming
explicitly that the state of a vertex is independent of the state of
its nearest neighbors.  As we will see, this assumption is not
correct, especially for low fire (infection) densities, when the
positions of different fires are in fact strongly correlated, leading
thus to a  breakdown of the HMF theory predictions (see
Sec.~\ref{sec:4}).

\section{Steady-state solution in general networks}
\label{sec:2}

Let us consider the long time properties of the FFM. It typically
corresponds to the steady-state calculated by setting $\dot
\rho^E_k=\dot \rho^T_k=\dot \rho^F_k=0\ \forall k$ in the HMF
Eqs.~\eqref{eq:mean_field}, which yields the algebraic equations
\begin{equation}\label{eq:stable_mean_field}
  \begin{cases}
    0=\rho^F_k -  p \rho^E_k\\ 
    0=p \rho^E_k -
    \rho^T_k \left[h k\sum_{k'}P(k'|k)\rho^F_{k'}+ g \right]\\ 
    0=\rho^T_k\left[h
      k\sum_{k'}P(k'|k)\rho^F_{k'}+g \right]
    - \rho^F_k 
  \end{cases}
\end{equation}
for the (now time-independent) variables $\rho^\al_k$. We look for
nontrivial steady-states, so we will be concerned in searching
solutions with $\rho^\al_k \neq 0.$

The analysis of Eqs.~(\ref{eq:stable_mean_field}) can be simplified by
noticing that the empty state plays the role of a rest state
(c.f. Appendix A), which the system enters and leaves with constant
rates. It can thus be factorized by writing its population density,
from the first equation in (\ref{eq:stable_mean_field}), as $\rho^E_k
= \rho^F_k/p$ and substituting it in the other two
equations. Therefore, introducing the rescaling factor
\begin{equation}
\eta=1+\frac{1}{p}
\end{equation}
and defining the new variable and parameter
\begin{equation}\label{eq:defNewVars}
\bar{\rho}^F_k \equiv \eta \rho^F_k, \qquad \ \ \bar{g}\equiv\eta g, 
\end{equation}
we can consider the simplified set of equations
\begin{equation}\label{eq:fixed_point}
\begin{cases}
  0=\bar{\rho}^F_k -
  \rho^T_k\left[hk\sum_{k'}P(k'|k)\bar{\rho}^F_{k'}+\bar{g} \right]\\
  1=\bar{\rho}^F_k + \rho^T_k
\end{cases}
\end{equation}
as characterizing the steady-state of the FFM in a general complex
network with a correlation pattern given by the conditional
probability $P(k'|k)$.

\begin{figure}[t]
\begin{center}
\includegraphics[width=0.18\textwidth]{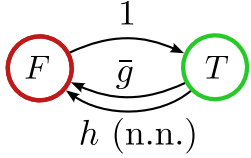}
\end{center}
\caption{Representation of the two states present in the SIS+g model,
  and of its dynamical rules. Its steady-state is directly related to
  that of the FFM.  (n.n. stands for a ``nearest neighbor''
  interaction).}
\label{fig:SIS+}
\end{figure}

From Eq.~(\ref{eq:fixed_point}), we can see that the steady-state of the FFM
can be in general mapped to the steady-state of an SIS model
\cite{Romualdo:2001} with a random source of infected individuals,
arising from isolated susceptibles with rate $\bar g$.  We can thus refer
to it as a SIS+g model, see Fig.~\ref{fig:SIS+}. In particular,
setting $g=0$, the FFM becomes the SIRS model, which is therefore
exactly mappable to the SIS model \cite{liu04:_spread}.

In order to solve the set of Eqs.~(\ref{eq:fixed_point}), one can
proceed to substitute its second equation 
into its first one, to obtain $\bar{\rho}^F_k$ as a function of
$\bar{\theta}_k \equiv \eta \theta_k$, namely
\begin{equation}\label{eq:f_from_theta}
  \bar{\rho}^F_k = \frac{h
    k\bar{\theta}_k+\bar{g}}{1+(hk\bar{\theta}_k+\bar{g})}.
\end{equation}
The equation is closed by expressing  $\bar{\theta}_k$
self-consistently as
\begin{equation}\label{eq:self_consistent}
  \bar{\theta}_k  = 
  \sum_{k'}P(k'|k)
  \frac{hk'\bar{\theta}_{k'}+\bar{g}}{1+(hk'\bar{\theta}_{k'}+\bar{g})}. 
\end{equation}
Solving this equation for $\bar{\theta}_k$ directly gives the
steady-state fire density, and thus the stationary solution of the
process. 

Prior to solving these equations, however, notice that
Eqs.~(\ref{eq:fixed_point}) imply that the solution for
$\bar{\rho}^F_k$ takes the functional form $\bar{\rho}^F_k(\bar{g},
h)$, and is now independent of $p$ (namely of $\eta$).  Therefore, we
can write down a scaling solution for the steady-state fire density in
the FFM in any network as
\begin{equation}
  \rho^F(\eta, g, h)=\frac{1}{\eta} \bar{\rho}^F(\eta g, h).
  \label{eq:8}
\end{equation}
This scaling solution implies that the factor $p$ (growth of new
trees) affects the model only by a rescaling of the fire density and of
the rate of spontaneous lightning. This justifies the fact that no
more than 2 out of the 4 initial parameters are relevant to the study
of the stationary state of the FFM within the HMF theory
approximation. For the particular case $g=0$, Eq.~(\ref{eq:8}) leads
to
\begin{equation}
  \rho^F(\eta, g=0, h)\equiv\frac{1}{\eta} F(h),
  \label{eq:9}
\end{equation}
that is, the fire density is a function of $h$, divided by $\eta$. In
the limit $p \to \infty$ ($\eta \to 1$), we recover the standard SIS
model. On the other hand, any finite $p$ will lead to a smaller value
of the fire density. The fact that an upper bound on $\rho^F$ can be
deduced from $\eta$ is in general valid for any $g$:
since $\bar{\rho}^F_k\leq1$, given that it is a probability, we have
that
\begin{equation}\label{eq:limit}
\rho^F \leq  \frac{1}{\eta} = \frac{p}{p+1}.
\end{equation}
This upper bound on $\rho^F$ has important consequences from a
numerical point of view. In fact, even in the regions of the parameter
space $(h, g)$ where a nonzero value of $\rho^F$ is expected, a very
small value of $p$ will lead to a correspondingly small fire density,
which can be difficult to measure unless in the limit of very large
network sizes.

The factorization of the empty state, the functional form of the fire
density and its upper bound in terms of $\eta$ are in fact quite
general features, that can be found in any dynamical system having
rest states, see Appendix A.

\subsection{Homogeneous networks}\label{Homogeneous}

Let us consider first the simplest situation of the FFM taking place
in a homogeneous network, in which the degree distribution is peaked
at an average degree $\avk$ and decays exponentially fast for $k\ll
\avk$ and $k\gg \avk$. We can approximate all vertices as having the
same degree $k = \avk$. In this case, we have $\rho^F_k \equiv \rho^F$
effectively independent of $k$, and also $\theta_k =
\rho^F$. Eq.~\eqref{eq:f_from_theta} thus takes the form
\begin{equation}\label{eq:quadratic}
  h\avk (\bar{\rho}^F)^2 + (1+\bar{g}-h\avk)\bar{\rho}^F - \bar{g} =
  0, 
\end{equation}
whose only positive solution is
\begin{equation}\label{eq:homogeneous}
  \bar{\rho}^F =
  \frac{h\avk-1-\bar{g}+\sqrt{-4h\avk+(1+\bar{g}+h\avk)^2}}{2h\avk}, 
\end{equation}
as already shown in Ref.~\cite{Christensen}.

For $g>0$ the fire density is strictly
positive. This results from the fact that the trees in the network can
always ignite themselves with some nonzero probability, therefore
always reviving the fire density. On the other hand, for $g=0$, the
fire density takes the form
\begin{equation}
  \rho^F = \frac{|1-h\avk|-(1-h\avk)}{2\eta h\avk}.
  \label{eq:6}
\end{equation}
which is equal to $0$ for $h < \avk^{-1}$ and positive otherwise. That
is, the FFM experiences an absorbing-state phase transition
\cite{marro99} at a critical value $h_c=\avk^{-1}$. For $h > h_c$, the
systems is in an active phase, in which the fire activity never stops,
taking the asymptotic form
\begin{equation}
  \rho^F \sim \frac{ h - h_c}{\eta}.
\end{equation}
On the other hand, for $h < h_c$, the system reaches an absorbing
state, in which fire always ends up disappearing by lack of
transmissibility.

\subsection{Phase transition for $g=0$ on complex networks}
\label{sec:2.5}

For networks with a general degree distribution $P(k)$ and general
correlation pattern $P(k'|k)$, the explicit solution of
Eqs.~(\ref{eq:f_from_theta}) and~(\ref{eq:self_consistent}) becomes a
quite difficult task. It is possible, however, to obtain information
for a general network in the particular case $g=0$.  Setting $g$ to
$0$ changes the forest fire model to a SIRS model, whose stationary
state can be related to the one of the SIS model by removal of the
rest state, see Eq.~(\ref{eq:fixed_point}). Therefore, in this
particular limit, the FFM exhibits an absorbing-state phase transition
between an active (burning, infected) phase and an absorbing
(fire-free, healthy) phase, located at the critical point
\cite{Epidemic_correlated_1}
\begin{equation}
h_c=\frac{1}{\Lambda_m},
\end{equation}
where $\Lambda_m$ is the largest eigenvalue of the connectivity matrix 
$C_{kk'}=kP(k'|k)$. Interestingly, this threshold is independent of
the rate $p$ of creation of new trees, which only affects the overall
density of fires, as expressed in Eq.~(\ref{eq:9}).

\section{Explicit solution for uncorrelated scale-free networks}
\label{sec:3}

In order to obtain explicit analytical results for the HMF equations of
the FFM, we restrict ourselves to the case of uncorrelated
networks. In this case, the conditional probability $P(k'|k)$ takes
the form $P(k'|k) = k' P(k') /\avk$ \cite{caldarelli2007sfn}.
$\bar{\theta}_k$ thus becomes independent of $k$:
\begin{equation}
  \bar{\theta}_k \equiv \bar{\theta} = \frac{1}{\avk} \sum_k k P(k)
  \bar{\rho}^F_k. 
\end{equation}
From Eq.~(\ref{eq:f_from_theta}), $\bar{\rho}^F_k$ is now an
algebraic function of $k$. For the interesting case of SF networks,
with a degree distribution $P(k) =
(\gamma+1)m^{\gamma+1}k^{-2-\gamma}$ in the continuous degree
approximation, where $m$ is the minimum degree present in the network,
Eq.\eqref{eq:self_consistent} reads, replacing summations by
integrals,
\begin{align}
  \bar{\theta} &= \gamma m^{\gamma}\int_m^\infty \frac{h\bar{\theta}
    k^{-\gamma}+\bar{g} k^{-1-\gamma}}{1+\eta(hk\bar{\theta}+\bar{g})}\\
  \begin{split}
    &= F\left(1, \gamma; \gamma+1;
      -\frac{1+\bar{g}}{mh\bar{\theta}}\right)\\ 
    &\ \ +\frac{\gamma \bar{g}}{(\gamma+1)mh\bar{\theta}}F\left(1,
      \gamma+1; \gamma+2; -\frac{1+\bar{g}}{mh\bar{\theta}}\right) 
\end{split}
\end{align}
where $F\left(a,b;c;z\right)$ is the Gauss hypergeometric function
\cite{Handbook}. Using the power series development of the
hypergeometric function \cite{Handbook} or the asymptotic expression
\begin{equation}\label{hypergeometric_developpement}
  F(1, \gamma; 1+\gamma; -z^{-1}) = \gamma\sum_{n=1}^\infty
  (-1)^n\frac{z^n}{n-\gamma} +
  \frac{\gamma\pi}{\sin(\gamma\pi)}z^\gamma ,
\end{equation}
valid for $\arg(z) < \pi$ and $\gamma \notin \mathbb{N}$, one finds the
self-consistent equation for $\bar{\theta}$ to be:
\begin{equation}\label{eq:self_theta}
\bar{\theta} = \frac{\bar{g}}{1+\bar{g}}+\frac1{1+\bar{g}}F\left(1,
    \gamma; \gamma+1; 
  -\frac{1+\bar{g}}{mh\bar{\theta}}\right) ,
\end{equation}
for all $\frac{1+\bar{g}}{mh\bar{\theta}}>0$. This is the final
equation we need to solve in order to find the steady state fire
density. Note that the condition of validity $\gamma \notin
\mathbb{N}$ is no more a restriction here, by analytical continuation
of the hypergeometric function.

Before proceeding, let us express more explicitly the dependence of
the fire density (what we are ultimately interested in) and the
probability for a neighbor to burn, $\bar \theta$. We can directly
calculate it using Eq.~\eqref{eq:f_from_theta} and a similar reasoning
as before, to obtain:
\begin{align}\label{eq:f_theta}
  \bar{\rho}^F &= \sum_k P(k) \bar{\rho}^F_k=
      (\gamma+1)m^{\gamma+1}\int_m^\infty
      \frac{hk\bar{\theta}+\bar{g}}{1+\eta(hk\bar{\theta}+\bar{g})} \nonumber\\   
      &=\frac{\bar{g}}{1+\bar{g}}+\frac1{1+\bar{g}}F\left(1, \gamma+1; \gamma+2;
    -\frac{1+\bar{g}}{mh\bar{\theta}}\right).
\end{align}
Now, using one of Gauss's relations for contiguous hypergeometric
functions \cite{Handbook}, namely
\begin{equation}
  (a-c)zF(a,b;c+1;z)+cF(a,b-1;c;z)=c(1-z)F(a,b;c;z)
\end{equation}
with $a=1$, $b=\gamma+1$ and $c=\gamma+1$, the
hypergeometric function can be re-expressed directly in terms of
$\bar{\theta}$ as obtained in Eq.~\eqref{eq:self_theta} to get
\begin{equation}\label{eq:rho_f}
  \bar{\rho}^F = \frac{\bar{g}}{1+\bar{g}} + \frac{\gamma+1}\gamma
    \frac{mh\bar{\theta}}{1+\bar{g}}(1-\bar{\theta}) 
\end{equation}
In a general SF network without correlations, the
fire density is thus a quadratic function of the probability
$\bar{\theta}$.

\subsection{Exact solution for $\gamma=1$}
\label{sec:3.1}
In the case $\gamma=1$ it is possible to solve exactly the
self-consistent equation Eq.~\eqref{eq:self_theta}, which takes the
form \cite{Handbook}
\begin{equation}
  \bar{\theta}=\frac{\bar{g}}{1+\bar{g}} +
  \frac{mh\bar{\theta}}{(1+\bar{g})^2}\ln\left(1+\frac{1+\bar{g}}{mh\bar{\theta}}\right)
  . 
\end{equation}
Introducing the new variable
$y=\bar{g}+\frac{\bar{g}(1+\bar{g})}{mh\bar{\theta}}$ the equation
becomes
\begin{equation}
  ye^y=\bar{g}e^{\bar{g}+\frac{(1+\bar{g})^2}{mh}}.
\end{equation}
Recognizing the solution of $ye^y=x$ as the W Lambert function,
$y=W(x)$ \cite{corless96}, the result follows:
\begin{equation}\label{eq:gamma1}
  \bar{\theta}=\frac{1+\bar{g}}{mh}\left[-1+\bar{g}^{-1}W
    \left(\bar{g}e^{\bar{g}+\frac{(1+\bar{g})^2}{mh}}\right)\right]^{-1}  
\end{equation}

The value of the fire density is then obtained by plugging this
expression into Eq.~\eqref{eq:rho_f}. Expanding the fire density at
first order for small $\bar{g}$ yields then
\begin{equation}\label{eq:dtdg}
  \begin{split}
    \bar{\rho}^F & \simeq
    \frac{2hm\left(e^{\frac1{hm}}-1\right)-2}{hm\left(e^{\frac1{hm}}-1\right)^2}
    + \left[3+\frac8{h^2m^2\left(e^{\frac1{hm}}-1\right)^3}\right.\\
    &\left.+\frac{8-10hm}{h^2m^2\left(e^{\frac1{hm}}-1\right)^2}-
      \frac{8-2hm}{hm\left(e^{\frac1{hm}}-1\right)}\right] \bar{g},
\end{split}
\end{equation}
where the first term corresponds, obviously, to the SIS result
\cite{Romualdo:2001}, recovered in the limit $g\to0$. The expansion in
terms of $h$ is
\begin{equation}\label{eq:dtdh}
  \bar{\rho}^F \simeq
  \frac{\bar{g}}{1+\bar{g}}+\frac{2\bar{g}m}{(1+\bar{g})^3}h,
\end{equation}
yielding a nonzero fire density (an infected steady-state) for any value of $h$ if $\bar g>0$.

\subsection{Asymptotic solution for $\gamma\neq1$}
\label{sec:3.2}

Let us now turn our attention to the behavior of the FFM in the
general case $\gamma\neq1$.  To do so, we will study the limit of low
fire density, namely $\bar{\theta} \ll 1$ and $\bar{g} \ll 1$.
In this limit, Eq.~\eqref{eq:rho_f} leads to  
\begin{equation}
  \bar{\rho}^F \sim \bar g + \frac{\gamma+1}{\gamma} m h \bar{\theta}, 
\end{equation}
so we need to develop the self-consistent equation for $\bar{\theta}$,
Eq.~\eqref{eq:self_theta}, up to the first most relevant terms, using
the asymptotic expansion Eq.~(\ref{hypergeometric_developpement}).  For
$\bar{g}=0$ we recover the known result $ \bar{\rho}^F \sim (h -
h_c)^\beta$ \cite{Endemic_states}, where
\begin{equation}
  h_c  = 
  \begin{cases}
    0 & \mathrm{for}\quad 0 < \gamma <1\\
    \frac{\gamma-1}{\gamma m}  & \mathrm{for}\quad  \gamma >1
  \end{cases},
\end{equation}
and
\begin{equation}
  \beta  = 
  \begin{cases}
    \frac{1}{1-\gamma} & \mathrm{for}\quad 0 < \gamma <1\\
    \frac{1}{\gamma-1} & \mathrm{for}\quad 1 < \gamma <2\\
    1  & \mathrm{for}\quad  \gamma >2
  \end{cases}.
\end{equation}

For $\bar{g} >0$, let us consider separately each possible value of
$\gamma$.

(1) $0 < \gamma < 1$:

In this case, the leading approximation for $\bar{\theta}$ is
\begin{equation}\label{eq:0gamma1}
  \bar{\theta} \simeq \frac{\bar{g}}{1+\bar{g}} +
  \frac{\gamma\pi}{(1+\bar{g})\sin(\gamma\pi)}
  \left(\frac{mh\bar{\theta}}{1+\bar{g}}\right)^\gamma,
\end{equation}
which is valid for $\frac{mh\bar{\theta}}{1+\bar{g}}\ll1$.  If $\bar g
\neq 0$, the solution of $\bar{\theta}$ depends in a nontrivial way on
both $\bar g$ and $h$ as 
\begin{equation}
  \bar{\theta} =
  \kappa\left(\frac{\bar{g}}{1+\bar{g}},
    \frac{\gamma\pi}{(1+\bar{g})\sin(\gamma\pi)}
    \left[\frac{mh}{1+\bar{g}}\right]^\gamma\right)  
\end{equation}
where the function $\kappa(a,b)$ is defined as the solution
of the implicit equation
$\kappa(a,b)=a+b\cdot\kappa(a,b)^\gamma$. In Appendix B we give an explicit
expression for $\kappa(a,b)$ in terms of $a$ and $b$.  For
small $\bar g$ and $h$, two different regimes can be isolated from the
development of the $\kappa$ function \eqref{eq:kappa}:
\begin{itemize}
\item{$h \ll \bar g^\frac{1-\gamma}\gamma$:} The most significant terms are now :
  \begin{align}
    \bar{\theta} &\simeq \frac{\bar{g}}{1+\bar{g}} +
    \frac{\gamma\pi}{(1+\bar{g})\sin(\gamma\pi)}
    \left(\frac{mh\bar{g}}{(1+\bar{g})^2}\right)^\gamma,
  \end{align}
  so that, to leading order in $h$, the fire density goes like:
  \begin{equation}
    \bar{\rho}^F \simeq \frac{\bar{g}}{1+\bar{g}} +
    \frac{\gamma+1}{\gamma}\frac{m\bar{g}}{(1+\bar{g})^3}h.
    \label{eq:5}
  \end{equation}
  The introduction of an infinitesimal $\bar{g}>0$ destroys again the
  absorbing-state phase transition, since both terms in
  Eq.~(\ref{eq:5}) are positive.
  
\item{$\bar g \ll h^\frac\gamma{1-\gamma}$:} In this case, the leading
  terms are:
  \begin{equation}
    \bar{\theta} \simeq
    \bar{\theta}_0+(1-\gamma)^{-1}\bar g ,
  \end{equation}
  where $\bar{\theta}_0=
  \left(\frac{\gamma\pi(mh)^\gamma}{\sin(\gamma\pi)}\right)^{\frac1{1-\gamma}}$. Thus,
  Eq.~\eqref{eq:rho_f} yields
  \begin{equation}
      \bar{\rho}^F \simeq
      \frac{\gamma+1}{\gamma}mh\theta^0(1-\bar{\theta}_0) +\left[1+\frac{(1+\gamma)mh}{\gamma(1-\gamma)}
        \left(1-2\theta_0\right)\right]\bar{g}. 
    \label{eq:4}
  \end{equation}
  One should notice that in order to derive this last expression we used
  Eq.~\eqref{eq:0gamma1}, which assumes
  $\frac{mh\bar{\theta}}{1+\bar{g}} \ll 1$, so not only do we need
  $\bar g$ to be small in order for this approximation to hold, but
  $h$ cannot be too large either.
  
\end{itemize}

(2) $\gamma > 1$: 

The self-consistent equation Eq.~(\ref{eq:self_theta}) can be
approximated in this regime by:
\begin{equation}\label{eq:g1}
  \bar{\theta} \simeq \frac{\bar{g}}{1+\bar{g}} +
  \frac\gamma{\gamma-1}\frac{mh\bar{\theta}}{(1+\bar{g})^2}.
\end{equation}
So we obtain:
\begin{equation}
  \bar{\theta} \simeq
  \frac{(\gamma-1)\bar{g}(1+\bar{g})}{(\gamma-1)(1+\bar{g})^2-\gamma
    mh}.
\end{equation}
and to first order, the fire density is given by
\begin{equation}
  \bar \rho^F \simeq \bar g\left(1 - \bar g + 2mh \right).
\end{equation}

To summarize, in all cases considered above the presence of a fire
lightning probability $g>0$ eradicates any phase transition in the
model, just like on homogeneous networks, Sec.~\ref{Homogeneous},
rendering a nonzero steady-state that grows, at lowest order, linearly
with $\bar{g}$, with a numerical prefactor which is a complex function
of $h$, depending on the particular value of $\gamma$ considered.

\section{Numerical Simulations}
\label{sec:4}

In order to check the analytic predictions developed in the previous
sections, we performed extensive numerical simulations of the FFM
on top of different network models, both homogeneous and
heterogeneous. Simulations were implemented using a sequential update
algorithm \cite{marro99}. Given a substrate network, each vertex is
first randomly initialized in one of the three possible states, empty,
tree, or fire. In the dynamics, every time step a vertex is chosen at
random and its state is updated applying the rules defined in
Sec.~\ref{sec:1} with given reaction rates $\ell$, $p$, $g$ and
$h$. In principle, to reproduce the exact dynamics a time step would
correspond to a time increment $\Delta t=1/N$ \cite{marro99}. To speed
up the arrival to the stationary state, we chose $\Delta t$ larger,
but still small enough so as to keep the process random. Typically,
the probabilities involved in the evolution are such that $\max(\ell
\Delta t,\ p\Delta t,\ g\Delta t,\ h\Delta t) \leq 0.1$. We have
checked that smaller probabilities do not affect the properties of the
steady state, but only slow down the transient to reach it. We can
recover the correct theoretical expressions from our HMF analysis just
by substituting the model parameters by the rescaled values $h/\ell$,
$g/ \ell$, $p /\ell$, and, correspondingly, $\eta = 1 + \ell / p$.

\subsection{Homogeneous networks: The Watts-Strogatz model} 
\label{nonMF}

As an example of a homogeneous network, we consider the small-world
model proposed by Watts and Strogatz (WS) \cite{WS}. Networks in this
model are generated as follows: The starting point is a ring with $N$
vertices, in which every vertex is symmetrically connected to its $2K$
nearest neighbors. Then, for every vertex, each edge connected to a
clockwise neighbor is rewired to a randomly chosen vertex with
probability $p_{rw}$, and kept with probability $1-p_{rw}$. This
procedure generates a graph with a degree distribution that decays
faster than exponentially for large k, and average degree $\avk =
2K$. We considered here WS networks with $p_{rw}$=1 and $K=3$.

\begin{figure}
  \begin{center}
    \includegraphics[width=0.9\columnwidth]{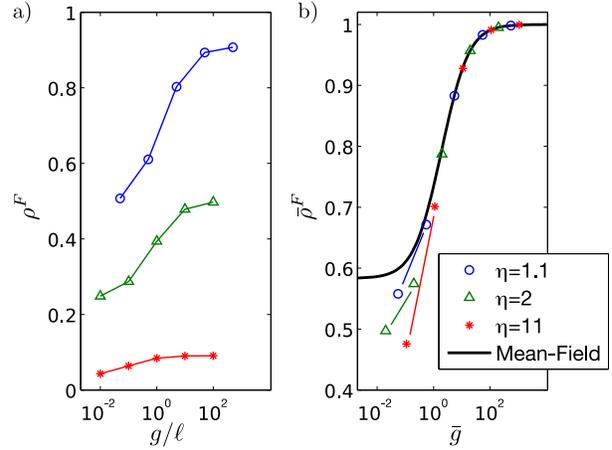}
  \end{center}
  \caption{Steady state fire density in the FFM on homogeneous WS
    networks of size $N=10^5$. (a) Raw data for different values of
    $\eta$ as a function of $g/\ell$, and fixed $h/\ell=0.4$. (b) Data
    collapse of the previous plots as given by Eq.~(\ref{eq:WS}).  The
    functions agree very well with the prediction by HFM theory for
    large values of the ratio $ \bar{g} \equiv \eta g / \ell$. The
    separation of the curves at low $\bar{g}$ indicates a possible
    non-mean-field regime.}
\label{fig:collapse}
\end{figure}

In Fig.~\ref{fig:collapse} we show the numerical results obtained in
WS homogeneous networks of size $N=10^5$, with fixed parameters $\ell
= 10^{-3}$ and $h/\ell=0.4$, and varying values of $p$ and $g$. We
chose in particular a large value of $h/\ell$, larger than $h_c/\ell =
\avk^{-1}=1/6$, in order to avoid possible problems in the vicinity of
the critical point for small values of $g$.  In this case, the
theoretical prediction for the fire density is given by
Eq.~(\ref{eq:homogeneous}), with the correct rescaling of parameters
\begin{equation}
  \rho^F =
  \frac{\frac{6 h}{\ell}-1-\eta \frac{g}{\ell}+\sqrt{-24\frac{h}{\ell}
      +(1+\eta \frac{g}{\ell}+6 \frac{h}{\ell})^2}}{12\frac{h}{\ell}\eta},  
  \label{eq:WS}
\end{equation}
indicating, as argued in Sec.~\ref{sec:2}, that $\eta \rho^F$ should
be a scaling function of $\eta g / \ell$.  As we can observe in
Fig.~\ref{fig:collapse}(b), the theoretical prediction is very well
satisfied by numerical data for large values of $ \eta g / \ell$,
collapsing all plots on the functional form of Eq.~ \eqref{eq:WS} when
the appropriate rescaling is performed. However, for values of $ \eta
g / \ell \leq 10$, we also observe a noticeable departure from the
mean-field prediction, which is more conspicuous for large values of
$\eta$. One could naively attribute this departure to a simple
numerical artifact: since we have in general that $\rho^F \sim
\eta^{-1}$, one could argue that, for fixed $h$, $g$, and $\ell$,
larger values of $\eta$ lead to ever smaller fire
densities. Therefore, for sufficiently large $\eta$, we could expect
such small fire density that its steady state determination will incur
in numerical resolution problems, unless extremely large network sizes
are considered. But here the minimal fire density we find is of the
order $\sim 0.04 \gg \frac{1}{N}=10^{-5}$ and so the number of burning
trees still represents a significant fraction of the total population.

\begin{figure}[t]
\begin{center}
\includegraphics[width=0.9\columnwidth]{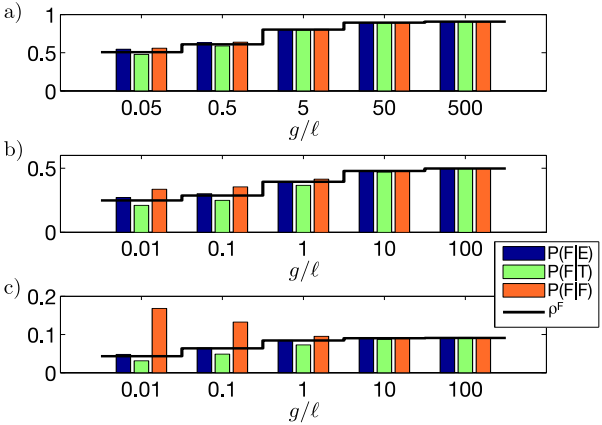}
\end{center}
\caption{Dependence of a vertex's state on his neighbors' states in WS
  networks of size $N=10^5$. The
  three graphs correspond to the three different curves presented in
  Fig.\ref{fig:collapse}: a) $\eta=1.1$, b) $\eta=2$, c)
  $\eta=11$. The red curve is the fire density $\rho^F$ in each
  case. In the absence of state correlations, we should have
  $P(F|E)=P(F|T)=P(F|F)=\rho^F$.}
\label{fig:statesCorrelations}
\end{figure}

A more thorough analysis, however, reveals that the actual reason of
this departure is the failure of the mean-field assumption of lack of
dynamical correlations between vertices discussed in
Sec.~\ref{mean-field} in the limit of small $ \eta g / \ell$ , and for
large $\eta$ (small $p /\ell$). In this case, the fire density is very
small, and therefore burning vertices are very likely nearest
neighbors of other fires, precisely those that originated them. This
fact introduces correlations between the state of nearest neighbors which
invalidate the whole mean-field approximation. We can check this
argument in a homogeneous network by comparing the probability
$P(\alpha)$ that any vertex is in state $\alpha$, with the
conditional probability $P(\alpha|\beta)$ that a vertex is in state
$\alpha$, provided it is nearest neighbor of a vertex in state
$\beta$. In Fig.~\ref{fig:statesCorrelations} we compare the numerical
values of $P(F) \equiv \rho^F$ with the corresponding conditional
probabilities $P(F|E)$, $P(F|T)$ and $P(F|F)$ in simulations performed
for different values of $\eta$, plotted as a function of $g/\ell$.
The figure shows that $P(F|F)$ is clearly larger that $\rho^F$ in the
areas of dissension: for small $g / \ell$, and especially for the
largest values of $\eta$. This results confirms the presence of strong
dynamical correlations between vertices, and hints towards the failure
of the HMF approximation in this region of the parameter space.

\subsection{Heterogeneous networks: The Barab\'{a}si-Albert model}  
\label{sec:4.2}

The Barab\'{a}si-Albert (BA) model is an algorithm to generate growing
SF networks with degree exponent $\gamma=1$, based on the
preferential attachment paradigm \cite{BA}. This model is defined as
follows: we start from a small number $m_0$ of vertices, and at each
time step, a new vertex is introduced, with $m$ edges that are
connected to old vertices $i$ with probability $\Pi(k_i) = k_i/ \sum_j
k_j$, where $k_i$ is the degree of the $i^{\text{th}}$ vertex. After iterating
this procedure a large number of times, we obtain a network composed
by $N$ vertices, minimum degree $m$, fixed average degree $\avk = 2m$,
degree distribution $P(k) = 2 m^2 k^{-3}$ and almost vanishing degree
correlations \cite{alexei02,barrat:_rate}. The simulations considered
here were perform with $m=2$ and a network size $N=10^6$.

In the case of the BA network, we can check the accuracy of the exact
HMF solution for $\gamma=1$, as given by Eqs.~(\ref{eq:rho_f})
and~(\ref{eq:gamma1}). This is a nontrivial function of two variables,
$g$ and $h$. Therefore, to check it in a simple way, we focused on the
small $g$ and $h$ regimes, in which expressions \eqref{eq:dtdg} and
\eqref{eq:dtdh} should provide a good approximation. In particular, we
computed the quantities $\partial \bar \rho^F / \partial \bar{g}$
and $\partial \bar\rho^F / \partial h$ which, for small $h/\ell$ and
$\bar{g}$ can be approximated by
\begin{equation}
\begin{split}
  \left.\frac{\partial \bar{\rho}^F}{\partial \bar{g}}\right|_{\bar g =0} &= 3+\frac8{h^2m^2\left(e^{\frac1{hm}}-1\right)^3}+\frac{8-10hm}{h^2m^2\left(e^{\frac1{hm}}-1\right)^2}\\
  &\ \ \ -\frac{8-2hm}{hm\left(e^{\frac1{hm}}-1\right)},\\
  \left.\frac{\partial \bar{\rho}^F}{ \partial h}\right|_{h=0} &=\frac{2\bar{g}m}{(1+\bar{g})^3}.
  \label{eq:7}
\end{split}
\end{equation}

To obtain those quantities numerically, we used computer simulations
to find the stationary fire density for several values of the variable
with respect to which we took the partial derivative. Those values
were chosen sufficiently small ($\bar g<10^{-3}$ and
$h/\ell<2\cdot10^{-3}$) so that the expected second order term is
negligible. We then checked that the points formed the expected
straight line, on which we measured the slope. The error on this slope
comes from statistical uncertainties on the measured fire density for
the points considered.

\begin{figure}
  \begin{center}
    \includegraphics[width=0.9\columnwidth]{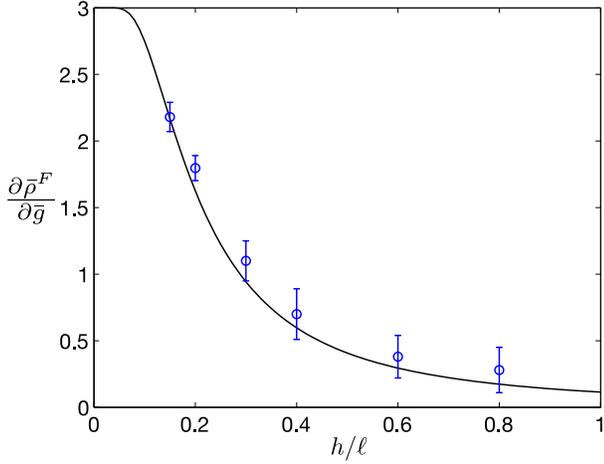}
  \end{center}
  \caption{Slope $\frac{d\bar\rho_f}{d\bar{g}}$ as a function of
    $h/\ell$ evaluated at $\bar{g}\leq 0.001$ on BA networks. The
    thick line is the mean field prediction for $\bar g=0$,
    Eq.(\ref{eq:7}).}
  \label{fig:4dfdg}
\end{figure}

\begin{figure}
\begin{center}
\includegraphics[width=0.9\columnwidth]{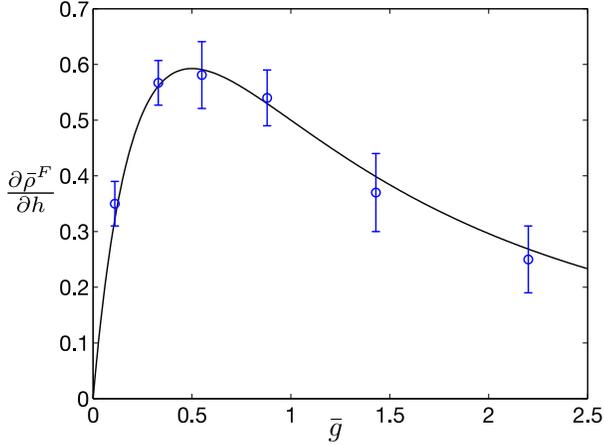}
\end{center}
\caption{Slope $\frac{d\bar \rho_f}{d\bar{h}}$ as a function of
  $\bar{g}$ evaluated at $h/\ell \leq 0.002$ on BA networks. The thick
  line is the mean field prediction for $h=0$, Eq.(\ref{eq:7}).}
\label{fig:4dfdh}
\end{figure}

Figs.~\ref{fig:4dfdg} and \ref{fig:4dfdh} show these quantities,
computed from numerical simulations of the FFM in BA networks
with $\ell=10^{-2}$. In order to ensure that we are within
the region of validity of the HMF prediction, we chose a small value
of $\eta=1.1$ ($p/\ell =10$). The good agreement observed between
numerical simulations and the theoretical HMF predictions in
Eqs.(\ref{eq:7}) confirms the  validity of the HMF analysis in this
parameter regime.

\subsection{Heterogeneous networks: The uncorrelated configuration model}
\label{sec:4.3}

To generate SF networks with an arbitrary degree exponent
$\gamma\neq1$, we used the uncorrelated configuration model
(UCM) \cite{UCM}. This model is defined as follows: We start from $N$
initially disconnected vertices. Each vertex $i$ is assigned a degree
$k_i$, extracted from the probability distribution $P(k)\sim
k^{-2-\gamma}$, subject to the constraints $m \leq k_i \leq N^{1/2}$
and $\sum_i k_i$ even. Finally, the actual network is constructed by
randomly connecting the vertices with $\sum_i k_i /2$ edges,
respecting the preassigned degrees and avoiding multiple and
self-connections. Using this algorithm, it is possible to create SF
networks whose average maximum degree (or cut-off) scales as $k_c(N)
\sim N^{1/2}$ for any degree exponent $\gamma$, and which are
completely uncorrelated \cite{Cut-offs}.  In the present simulations
we chose a minimum degree $m=2$, $\ell=10^{-2}$ and sizes up to
$N=10^6$.

\begin{figure}
\begin{center}
  \includegraphics[width=0.9\columnwidth]{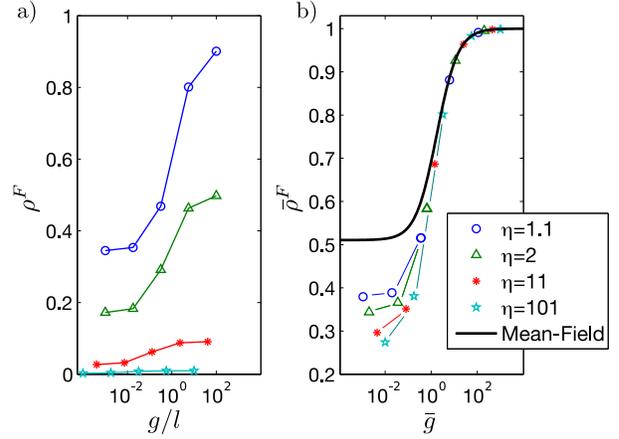}
\end{center}
\caption{Steady state fire density in the FFM on heterogeneous UCM
  networks with $\gamma=1/2$ and size N = $10^6$. (a) Raw data for
  different values of $\eta$ as a function of $g/\ell$, and fixed
  $h/\ell=0.4$.  (b) Data collapse of the previous plots as given in
  Eq.(\ref{eq:8}).  The full line is a numerical solution to
  Eq.~\eqref{eq:self_theta} plugged into \eqref{eq:rho_f} with the
  appropriate parameters ($m=2$, $\gamma=1/2$, $h/\ell=0.4$). For
  small $ \bar{g} \equiv \eta g / \ell$, deviations from the mean
  field theory are observed.}
\label{fig:collapseUCM}
\end{figure}

In Fig.\ref{fig:collapseUCM} we check the scaling of the fire
density $\eta \rho^F$ as a function of $\eta g / \ell$ in UCM networks
with degree exponent $\gamma=1/2$.  As in the case of the homogeneous
WS networks, the data collapse is very good for large values of $\eta
g / \ell$, fitting perfectly the theoretical prediction (full line)
obtained by a numerical resolution of Eqs.~\eqref{eq:self_theta} and
\eqref{eq:rho_f}. Again, deviations from the HMF prediction are
observed for small $\eta g / \ell$, which must be attributed to
the presence of strong dynamical correlations between vertices, which
invalidate the HMF approximation.

\begin{figure}[t]
\begin{center}
  \includegraphics[width=0.9\columnwidth]{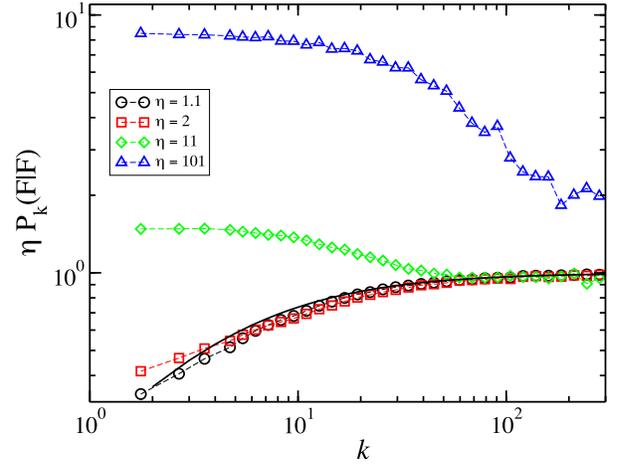}
\end{center}
\caption{Dependence of a vertex's state on his neighbors' states, as a
  function of the degree in UCM networks with $\gamma=\frac{1}{2}$ and
  size $N=10^6$. The thick line corresponds to the theoretical
  prediction $\bar{\rho}^F_k$, computed numerically from
  Eqs.~\eqref{eq:self_theta} and \eqref{eq:f_from_theta}.  Data for
  $h/\ell=0.4$, $\bar g=0.01$.}
\label{fig:statesCorrelationsUCM}
\end{figure}

The SF nature of this network model allows to explore the role of the
degree in the establishment of dynamical correlations at small values
of $\eta g / \ell$. In order to do so, we concentrate in this case on
the conditional probabilities $P_k(\alpha|\beta)$ that a vertex of
degree $k$ be in state $\alpha$, provided it is nearest neighbor of a
vertex in state $\beta$. In absence of dynamical correlations, we
should expect $P_k(\alpha|\beta) \equiv \rho^\al_k$ in the steady-state
of the dynamics. In Fig.~\ref{fig:statesCorrelationsUCM} we show the
theoretical value $\bar{\rho}^F_k$ for $\bar g=10^{-2}$, calculated
from Eqs.\eqref{eq:self_theta} and \eqref{eq:f_from_theta}, compared
with the conditional rescaled probability $\eta P_k(F|F)$, evaluated
from numerical simulations and plotted as a function of $k$ for
different values of $\eta$. In the absence of dynamical correlations,
we should observe the plots of $\eta P_k(F|F)$ to collapse onto the
theoretical curve $\bar{\rho}^F_k$ for the different values of
$\eta$. While this scenario is correct for $\eta \leq 2$, we observe
very strong deviations from the HMF prediction for large $\eta$,
signaling the non-mean-field behavior of the FFM in SF networks. In
particular, the conditional probability $P_k(F|F)$ turns out to be
larger than the HMF average value $\rho^F_k =
\eta^{-1}\bar{\rho}^F_k$, the difference increasing for small degree values.
The discrepancy between $P_k(F|F)$ and the average fire
density, also observed in homogeneous networks (see
Fig.~\ref{fig:statesCorrelationsUCM}), is again due to the clustering
at low fire densities of burning vertices in connected regions of the
network.

\section{Conclusion}
\label{concl}

In this paper we have presented a detailed analytical study of the
forest fire model (FFM) in complex networks. From the perspective of
the modelization of epidemic spreading processes, the FFM represents a
generalization of several well-known epidemic models previously
studied, which can be captured within the formalism of the FFM by the
appropriate selection of representative parameters. Applying the now
established HMF theory formalism, we have derived a set of rate
equations in continuous time that represent the dynamics of this
model. Focusing in the long term steady state behavior, we have
defined a set of algebraic equations, whose analysis allows to discuss
the role of the different parameters in the model. Thus, the rate at
which empty sites become trees (the recovery rate of infected
individuals) $p$, turns out to be absorbed in a rescaling of the fire
density (density of infected individuals) and of the spontaneous
ignition (spontaneous infection) rate, yielding in this way a fire
density that is inversely proportional to $1/p$. In the case of
homogeneous networks, we recover the results previously obtained in
the mean-field analysis of the FFM \cite{Christensen}. In the case of
heterogeneous SF networks, on the other hand, we have been able to
provide exact explicit expressions for the fire density for a degree
exponent $\gamma=1$, and approximate expressions for $\gamma<1$, valid
for the cases of $h$ or $g$ very small.

A comparison of these theoretical predictions with large scale
simulations in homogeneous and heterogeneous networks shows the HMF
theory to provide a correct description of the steady state of the FFM
for large $p$ (small $\eta$) and $g$, a regime in which the average
fire density is sufficiently large. For small $p$ (large $\eta$) and
$g$, on the other hand, numerical simulations indicate the breakdown
of HMF theory. The origin of this failure can be traced back to the
build up of dynamical correlations between nearest neighbor vertices,
correlations which in fact are expected to appear, due to the fact
that fires accumulate with large probability in connected clusters,
and are therefore not homogeneously distributed over the network as
assumed by mean-field approaches. 

Apart from providing new insights
into the behavior of a dynamical model relevant in epidemiological
modeling, our results indicate a possible path to the understanding of
the failure of HMF theory observed in other kinds of non-equilibrium
processes in complex networks \cite{castellano06:_non_mean}.

\begin{acknowledgement}
  R.P.-S.  acknowledges financial support from the Spanish MEC (FEDER)
  under Project No. FIS2007-66485-C02-01, and additional financial
  support through ICREA Academia, funded by the Generalitat de
  Catalunya.  The authors thank M. Moret for making available to
  us the Bionics cluster, and C. Castellano for helpful comments and
  discussions. 
\end{acknowledgement}

\vspace*{1cm}

\section*{Appendix A: Rest states play no role in steady-state
  solutions}
\label{sec:2.1}

\begin{figure}[t]
\begin{center}
  \includegraphics[width=0.5\columnwidth]{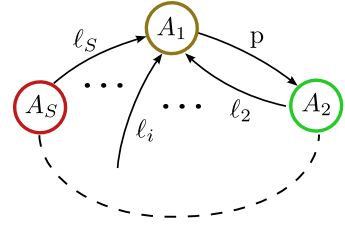}
\end{center}
\caption{General dynamical process with a rest state $A_1$.}
\label{fig:descanso}
\end{figure}

Let us consider a general dynamical process taking place on $S$ states
where $A_1$ is what we will call here a ``rest state'': every site
$A_i$ can evolve to $A_1$ only with some fixed rate $\ell_i$ and $A_1$
evolves to $A_2$ with some other fixed rate $p$, see
Fig. \ref{fig:descanso}.  The dynamical equations of this process take
the general form:
\begin{equation}\label{eq:generalMF}
  \begin{cases}
    \dot a_1(t)&=\sum_{i=2}^S\ell_i a_i(t) - p a_1(t)\\
    \dot a_2(t)&=p a_1(t) + f_2[a_2(t), ..., a_S(t)] - \ell_2 a_2(t)\\
    &\ \vdots\\
    \dot a_i(t)&=f_i[a_2(t), ..., a_S(t)] - \ell_i a_i(t)\\
    &\ \vdots\\
    \dot a_S(t)&=f_S[a_2(t), ..., a_S(t)] - \ell_S a_S(t) \\
    1&=\sum_{i=1}^S a_i
  \end{cases},
\end{equation}
for some functions $f_i[a_2(t), ..., a_S(t)]$, where $a_i(t)$ is the
probability of finding the system in state $i$ at time $t$.  Trying to
solve this system for the stationary state by setting $\dot a_i(t) =
0\ \forall i$, one can get rid of the $a_1$ variable by substituting
its value from the first equation into the other ones, namely
$a_1=\sum_i\ell_i/p\ a_i$. Then the last $S-1$ equations form a set of
$S-1$ independent equations with only $S-1$ unknowns:
\begin{equation}\label{eq:Sm1}
  \begin{cases}
    0&=\sum_{i=2}^S\ell_i a_i + f_2(a_2, ..., a_S) - \ell_2 a_2(t)\\
    &\ \vdots\\
    0&=f_i(a_2, ..., a_S) - \ell_i a_i(t)\\
    &\ \vdots\\
    0&=f_S(a_2, ..., a_S)-\ell a_S\\
    1&=\sum_{i=2}^S a_i + \sum_{i=2}^S\ell_i/p\  a_i
  \end{cases}.
\end{equation}

Now if the introduced functions can be written as
\begin{equation}
  f_i(a_2, ..., a_S)=\sum_{n_2,\ldots,n_S} c_i^{n_2,\ldots,n_S} \cdot
  a_2^{n_2}\cdot\ldots\cdot a_S^{n_S}, 
\end{equation}
then, performing the change of variable
\begin{equation}
  \bar{a}_i=\eta_i a_i,\
  \bar{c}_i^{n_2,\ldots,n_S}=c_i^{n_2,\ldots,n_S}\prod_j\eta_j^{-n_j},\
  \bar\ell_i = \ell_i\eta_i^{-1}\ \forall i 
\end{equation}
keeps all the first equalities in \eqref{eq:Sm1} true for any value of
the constants $\eta_i$. In particular, choosing
\begin{equation}
\eta_i=1+\ell_i / p,
\end{equation}
then all equalities are verified and the last one becomes
$\sum_{i=2}^S\bar{a}_i=1$.

In other words, the new variables $\bar{a}_i$, together with
Eqs.~\eqref{eq:Sm1} above, describe the stationary solution of a
dynamical system with analogous dynamics as the first one
\eqref{eq:generalMF}, up to a constant for some parameters, and
without the rest state. Finding the stationary solution of the
original $S$-states system is thus equivalent to finding the one for the
new system with $S-1$ states.

The new model has one parameter less, $p$, which has been absorbed in
the redefinition of all other parameters and variables of the new
system. So the stationary state of the first model is somehow
independent on $p$. This can be stated as:
\begin{equation}
\begin{split}
  \eta_i &a_i(\{\eta_j, \prod_k\eta_k^{n_k}\bar
  c_j^{n_2,\ldots,n_S},\eta_j\bar\ell_j\}_j)\\ 
  &=\bar a_i(\{\bar c_j^{n_2,\ldots,n_S},\bar\ell_j\}_j)\text{
    Indep. of }p\ \forall i
\end{split}
\end{equation}

Moreover, an upper bound for all densities $a_i$ can be deduced
directly from this argument: Since $\bar{a}_i$ is the probability to
be in state $i$ in the new system, then
\begin{equation}
\bar{a}_i <1 \Rightarrow a_i<\eta_i^{-1}.
\end{equation}

Notice that in the main text we chose the time scale so that
$\bar\ell=1$. This has the effect of applying the $\eta$ factor to
different coefficients (c.f. Eq. \eqref{eq:defNewVars}).

\section*{Appendix B: An explicit solution to a transcendental
  equation}
\label{explicit_transcendental}
\label{app:1}
Let's consider the equation
\begin{equation}\label{eq:trans}
  \kappa = a + b\kappa^\gamma
\end{equation}
for $\kappa$, where $a,\ b\in \mathbb{R}_+$ are given and $0<\gamma<1$
is fixed. This equation was already considered back in 1772 by Lambert
\cite{Lambert1772}.

If $a\neq0$ we can introduce the variable $v=\kappa/a$ and see that
this new variable depends on a unique variable $\beta=ba^{\gamma-1}$:
\begin{equation}
  v=1+\beta v^\gamma.
\end{equation}
Similarly, if $b\neq0$, the new variable $u=\kappa
b^{\frac1{\gamma-1}}$ only depends on $\alpha =
ab^{\frac{1}{\gamma-1}}=\beta^{\frac1{\gamma-1}}$:
\begin{equation}
u=\alpha+u^\gamma.
\end{equation}

So if we know how to express explicitly $v(\beta)$ or $u(\alpha)$, we
can conversely do so with $\kappa(a,b)$. To achieve this, we use the
Lagrange inversion theorem \cite{LagrangeInversion} which states that
the reciprocal function $x(y)$ of $y=f(x)$ can be expressed around
$y_0=f(x_0)$ as
\begin{equation}
x=x_0 + \sum_{n=1}^\infty c_n (y-y_0)^n
\end{equation}
with $c_n =
\left. \frac{d^{n-1}}{dx^{n-1}}\left(\frac{x-x_0}{f(x)-y_0}\right)^n
\right|_{x=x_0}\frac1{n!}$, if $f'(x_0)\neq0$.

(1) Case $a \neq 0$

In this case we have $\beta(v) = v^{-\gamma}(v-1)$ and so around
$v_0=1$ and $\beta_0=0$, $\beta'(1)\neq0$ and we can calculate:
\begin{align}
  n! c_n &=
  \left. \frac{d^{n-1}}{dv^{n-1}}\left(\frac{v-v_0}{\beta(v)-\beta_0}\right)^n
  \right|_{v=v_0}\\ 
  &= \frac{\Gamma(n\gamma+1)}{\Gamma(n(\gamma-1)+2)}.
\end{align}
Thus we can write
\begin{equation}\label{eq:vsol}
  v(\beta) = 1+\sum_{n=1}^\infty
  \frac{\Gamma(n\gamma+1)}{\Gamma(n(\gamma-1)+2)\Gamma(n+1)}\beta^n .
\end{equation}
This formulation of the function $v(\beta)$ is true as long as the
series converges. To know what interval it corresponds to, we
calculate the convergence radius $\beta_c$ following its definition:
\begin{align}
  \beta_c^{-1} &= \limsup_{n\to\infty}\sqrt[n]{|c_n|}\\
  &=
  \limsup_{n\to\infty}\sqrt[n]{\left|\sin(n\pi(1-\gamma))
      \frac{\Gamma(n\gamma+1)\Gamma(n(1-\gamma)-1)}{\Gamma(n+1)}\right|}\\  
  &= \lim_{n\to\infty}\sqrt[n]{
    \frac{\Gamma(n\gamma+1)\Gamma(n(1-\gamma)-1)}{\Gamma(n+1)}},
\end{align}
where we have used the Euler reflection formula to avoid an undefined
valued of the Gamma function \cite{havil03}. Using the Stirling
formula $\Gamma(z)=\sqrt{\frac{w\pi}z}\left(\frac
  ze\right)^z(1+\mathcal{O}(\frac1z))$ and
$\lim_{n\to\infty}\sqrt[n]{n}=1$ we find
\begin{equation}
\beta_c^{-1}=\gamma^\gamma (1-\gamma)^{1-\gamma}.
\end{equation}
So Eq.~\eqref{eq:vsol} can be used to express $\kappa$ explicitly, but
not for any value of its arguments.

(2) Case $b\neq 0$:

Let's find now a series representation of function $u(\alpha)$. To
simplify the calculation we introduce in this case $w=u^{1-\gamma}$ so
that 
\begin{equation}
  w^{\frac1{1-\gamma}}=\alpha+w^{\frac\gamma{1-\gamma}}.
\end{equation}
We can now directly apply the same steps as before to find a power
series expression for the reciprocal function of
$\alpha(w)=w^{\frac\gamma{1-\gamma}}(w-1)$ around $w_0=1$, $\alpha=0$,
since $\alpha'(1)\neq0$:
\begin{align}
  n! c_n &= \left. \frac{d^{n-1}}{dw^{n-1}}\left(\frac{w-w_0}
      {\alpha(w)-\alpha_0}\right)^n\right|_{w=w_0}\\ 
  &= \frac{\Gamma(-\frac{n\gamma}{1-\gamma}+1)}{\Gamma(-\frac
    n{1-\gamma}+2)}. 
\end{align}
Thus we have
\begin{equation}\label{eq:usol}
  u(\alpha) = \left[1+\sum_{n=1}^\infty
  \frac{\Gamma(-\frac{n\gamma}{1-\gamma}+1)}
  {\Gamma(-\frac{n}{1-\gamma}+2)\Gamma(n+1)}\alpha^n\right]^{\frac1{1-\gamma}}.  
\end{equation}

Again we can calculate the convergence radius and see that it is
$\alpha_c=(1-\gamma)\gamma^{\frac\gamma{1-\gamma}}=\beta_c^\frac1{1-\gamma}$. In
other words, we found a complete description of the solution
$\kappa(a,b)$ of Eq.~\eqref{eq:trans}: depending on whether the value
of $\beta$ is smaller or greater than $\beta_c$, development
\eqref{eq:vsol} or \eqref{eq:usol} converges and gives an explicit
value for $\kappa(a,b)$:

\begin{equation}\label{eq:kappa}
  \kappa(a,b) =
  \begin{cases}
    a\sum\limits_{n=0}^\infty\frac{\Gamma(n\gamma+1)}
    {\Gamma(n(\gamma-1)+2)\Gamma(n+1)}\left(ba^{\gamma-1}\right)^n 
    \\ 
    \ \ \ \ \ \ \ \ \ \ \ \ \ \ \ \ \ \ \ \ \ \ \ \ \ \ \ \ \ \ \ \ \
    \ \ \ \ \ \ \ \text{if } \beta < \beta_c\\ 
    \\
    \left[b\left(1-\sum\limits_{n=1}^\infty(-1)^{n}\frac{\Gamma\left(\frac
            n{1-\gamma}-1\right)}{\Gamma\left(\frac{n\gamma}
            {1-\gamma}\right)\Gamma(n+1)}\left(ab^{\frac1{\gamma-1}}
        \right)^{n}\right)\right]^{\frac1{1-\gamma}}   
    \\ 
    \ \ \ \ \ \ \ \ \ \ \ \ \ \ \ \ \ \ \ \ \ \ \ \ \ \ \ \ \ \ \ \ \
    \ \ \ \ \ \ \ \text{if } \beta > \beta_c 
  \end{cases}
\end{equation}
with
\begin{equation}\label{betac}
  \beta_c=\gamma^{-\gamma}(1-\gamma)^{-(1-\gamma)}\ \in\ ]1,2[.
\end{equation}
Since $\kappa$ depends smoothly on $a$ and $b$, as can be seen from
Eq.~\eqref{eq:trans}, the case $\beta=\beta_c$ is deducible as the
limit of any of the two series.

This expression gives the value of the function $\kappa(a,b)$ whenever
either $a$ or $b$ is different from zero. The solution in the case
$a=b=0$ is trivially $x=0$.


\end{document}